\documentclass[fleqn,dvips]{article}
\usepackage{amsfonts,amssymb,amsmath,graphics}
\usepackage{graphicx}
\usepackage{floatflt}
\usepackage{color,colordvi}
\usepackage[english]{babel}
\newcommand*{\dis}{\displaystyle}
\newcommand*{\bs}{\boldsymbol}
\newcommand*{\intl}{\int\limits}
\newcommand*{\suml}{\sum\limits}
\newcommand*{\prodl}{\prod\limits}

\title{\bf Collision Thermalization of Nucleons in Relativistic
       Heavy-Ion Collisions }
\author{ D. Anchishkin$^a$, A. Muskeyev$^b$, S. Yezhov$^b$
\\ $^a$ {\small \it Bogolyubov Institute for Theoretical Physics,
             03680 Kiev, Ukraine}
\\$^b$ {\small \it Taras Shevchenko Kiev National University,
03022 Kiev, Ukraine } }

\date{}
\begin{document}

\maketitle

%%%%%%%%%%%%%%%%%%%%%%%%%%%%%%%%%%%%%%%%%%%%%%%%%%%%%%%%%%%%%%%%%%%%%%

\begin{abstract}
We consider a possible mechanism of thermalization of nucleons in
relativistic heavy-ion collisions. Our model belongs, to a certain
degree, to the transport ones; we investigate the evolution of the
system created in nucleus-nucleus collision, but we parametrize this
development by the number of collisions of every particle during
evolution rather than by the time variable. We based on the
assumption that the nucleon momentum transfer after several
nucleon-nucleon (-hadron) elastic and inelastic collisions becomes a
random quantity driven by a proper distribution. This randomization
results in a smearing of the nucleon momenta about their initial
values and, as a consequence, in their partial isotropization and
thermalization. The trial evaluation is made in the framework of a
toy model. We show that the proposed scheme can be used for
extraction of the physical information from experimental data on
nucleon rapidity distribution.
\end{abstract}

%%%%%%%%%%%%%%%%%%%%%%%%%%%%%%%%%%%%%%%%%%%%%%%%%%%%%%%%%%%%%%%%%%%%%%
\section{Introduction}
%%%%%%%%%%%%%%%%%%%%%%%%%%%%%%%%%%%%%%%%%%%%%%%%%%%%%%%%%%%%%%%%%%%%%%

The problem of isotropization and thermalization in the course of
collisions between heavy relativistic ions attracts much
attention, because the application of thermodynamic models is one
of the basic phenomenological approaches to the description of
experimental data. Moreover, the assumption about a local
thermodynamic equilibrium, along with other factors, is
successfully used in various domains of high-energy physics.
Meanwhile, many of the questions concerning this issue remain open
for discussion \cite{Mueller 2002}-\cite{Kharzeev 2007}.

The main goal of the investigations of the collisions of
relativistic nuclei is the extraction of a physical information about
nuclear matter and its constituents. It is a matter of fact that
we can get know more about quarks and gluons (constituents) just
under extreme conditions, i.e. at high densities and temperatures.
During last two decades, one of the celebrated tools on the way of
the theoretical investigations of this subject was relativistic
hydrodynamics which started to be applied to elementary particle
physics by the famous Landau's paper \cite{landau-1953}.

Applying relativistic hydrodynamics, one can partially describe
experimental data and get know that the matter created in
relativistic nucleus-nucleus collisions can be regarded on some
stage of evolution as a continuous one, i.e. as a liquid. Moreover,
as was discovered in Brookheaven National Laboratory, it can be
regarded even as a perfect fluid
\cite{Gyulassy-0403032,Kapusta-0705.1277} which consistent with a
description of the created quark-gluon plasma (QGP).
Unfortunately, the important physical information is hidden
in sophisticated numerical codes which solve the Euler hydrodynamic
equations of motion.

In the present paper, we propose an approach to description of
relativistic heavy-ion collisions which allows one to extract the
physical information from experimental data on the basis of
a transparent analytical model.

%%%%%%%%%%%%%%%%%%%%%%%%%%%%%%%%%%%%%%%%%%%%%%%%%%%%%%%%%%%%%%%%%%%%%%
\section{The Model}
%%%%%%%%%%%%%%%%%%%%%%%%%%%%%%%%%%%%%%%%%%%%%%%%%%%%%%%%%%%%%%%%%%%%%%

Our model is aimed at the extraction of the physical information
from the nucleon spectra for such collision energies when the
number of created nucleon-antinucleon pairs is much less than the
number of net nucleons. This means that the model can be applied
at AGS and low SPS energies.

1. We separate all nucleons in the final state, i.e. after
freeze-out, into two groups in accordance with their origination:
a) the first group consists of net nucleons that went through just
hadron reactions;
b) the second group includes nucleons which were created in the
collective processes, for instance, during hadronization of the
QGP. In accordance with this, the nucleon momentum spectrum can be
represented as a sum of two different contributions:
\begin{equation}
\frac {d N}{d^3p} = \left( \frac {d N}{d^3p}
\right)_{\rm hadron} + \left( \frac {d N}{d^3p} \right)_{\rm
QGP} \,. \label{model1}
\end{equation}
In turn, the total number of registered  nucleons
equals $N_{\rm total} = N_{\rm hadron} + N_{\rm QGP}$. If this
separation can be done, then we can define the ``nucleon power''
of the created QGP as $P_{\rm QGP}^{(N)}=N_{\rm QGP}/N_{\rm
total}$. In our further investigation, we will deal mainly with the
nucleons from the first group which come to the final state after
a chain of sequential nucleon-nucleon (-hadron) elastic and
inelastic collisions.

2. The collision number for every nucleon (hadron) is finite
because the lifetime of the fireball is limited. To determine the
maximal number of collisions, $M_{\rm max}$, in a particular
experiment we use the results of UrQMD simulations
\cite{urqmd1,urqmd2}.

3. Because the colliding nuclei are the spatially restricted
many-nucleon systems, the different nucleons experience
different collision numbers: it is intuitively clear that the
collision histories of the inner and surface nucleons will be
different. That is why, we partition all amount of nucleons of the
first group into different ensembles in accordance with a number
of collisions before freeze-out. Then the nucleons from every ensemble
give their own contribution to the total nucleon spectrum. If we
denote the number of particles in a particular ensemble
where the particles experienced $M$ collisions by $C(M)$, then, in
correspondence to Eq. (\ref{model1}), we can write the total
nucleon spectrum as
\begin{equation}
\frac{d N}{d^{\,3} p}
= \sum_{M=1}^{M_{\rm max}} C(M) D_M(\bs p) +
C_{\rm therm} D_{\rm therm}(\bs p) \, ,
\label{model2}
\end{equation}
where $D_M(\bs p)$ is the spectrum (normalized to unity) of the
particles in the $M$-th ensemble. The last term on the r.h.s. of
(\ref{model2}) corresponds to the possible contribution from the
totally thermalized source which we associate with QGP. Here,
$C_{\rm therm}$ is the number of nucleons from the second group
which are created during the hadronization, and $D_{\rm therm}(\bs p)$
is the thermal distribution normalized to unity.

\begin{figure}
\includegraphics[width=0.48\textwidth]{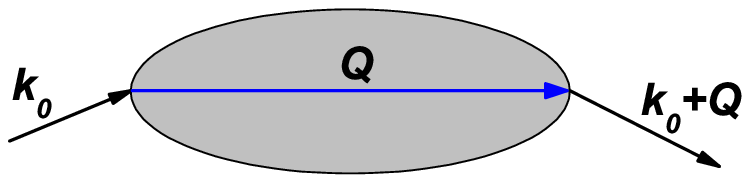}
\hfill
\includegraphics[width=0.49\textwidth]{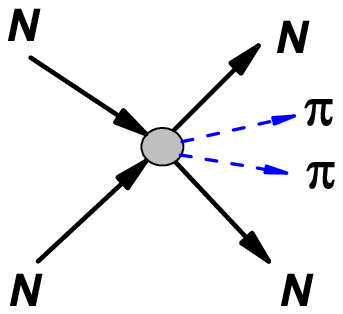}
\\
\parbox[t]{0.48\textwidth}{\caption{
Transformation of the initial nucleon momentum, $\bs k_0$, as a
result of $M$ collisions; ${\bs Q}=\sum_{m=1}^M{\bs q}^{(m)}$ is
the total momentum transfer after $M$ effective collisions, where
$\bs q^{(m)}$ is the momentum transfer in the $m$-th collision.}
\label{fig:totalmt}}
\hfill
\parbox[t]{0.48\textwidth}{\caption{
Example of inelastic collision of two nucleons with creation of
two $\pi$-mesons, $N+N \to N+N+\pi + \pi$. \label{fig:inelastic}
}}
\end{figure}

Consider successive variations of the momentum of the $n$-th
nucleon from nucleus $A$ which moves along the collision axis
from left to right toward nucleus $B$. Every $m$-th collision induces the momentum
transfer, ${\boldsymbol{q}} _n^{(m)}$, for the $n$-th nucleon.
So that, after $M$ collisions, the nucleon acquires the momentum
${\boldsymbol{k}}_n$:
\begin{equation}
{\bs k}_{0} \quad \rightarrow \quad {\bs k}_{0}+{\bs q}_{n}^{(1)}
\quad \rightarrow  \quad {\bs k}_{0}+{\bs q}_{n}^{(1)} +
{\bs q}_{n}^{(2)} \quad \rightarrow \quad \cdots  \quad \rightarrow
\quad {\bs k}_{0} + {\bs Q}_n \ \equiv \ {\bs k}_{n} \,,
\label{scattering-chain}
\end{equation}
where $\bs Q_n = \sum\limits_{m=1}^M \bs q_n^{(m)}$ is the total
momentum transfer finally obtained by the $n$-th nucleon (see
Fig.~\ref{fig:totalmt}).

%%%%%%%%%%%%%%%%%%%%%%%%%%%%%%%%%%%%%%%%%%%%%%%%%%%%%%%%%%%%%%%%%%%%%%
%%%%%%%%%%%%%%%%%%%%%%%%%%%%%%%%%%%%%%%%%%%%%%%%%%%%%%%%%%%%%%%%%%%%%%

Let us assume for the moment that the elastic scattering gives the
main contribution to the two-nucleon collision amplitude.
The initial momentum of every nucleon in nucleus $A$ is
$\boldsymbol{k}_a=\boldsymbol{k}_0=(0,0,k_{0z})$, while the
initial momentum of every nucleon in nucleus $B$ is
$\boldsymbol{k}_b=-\boldsymbol{k}_0=(0,0,-k_{0z})$.
The energy and momentum are conserved in every separate collision of
two particles,
$ \omega({\bs k}_a) + \omega({\bs k}_b) = \omega({\bs p}_a) +
\omega({\bs p}_b), ~~ \bs k_a + \bs k_b  =  \bs p_a + \bs p_b$,
where $\boldsymbol{p}_a$ and $\boldsymbol{p}_{b}$ are the momenta of
the particles after the collision.
We assume that the particles are on the mass shell,
so that $\omega (\boldsymbol{k})=\sqrt{m^2+\boldsymbol{k}^2}$ (the
system of units $\hbar =c=1$ is adopted).
Determining the six unknown quantities, $\bs p_a$ and $\bs p_b$, from
four equations is straightforward, but two quantities,
e.g. $(\boldsymbol{p}_a)_x$ and $(\boldsymbol{p}_b)_x$, remain
uncertain and can be considered as such which accept random values
driven by the scattering probability.
So, after the first collision, one component of the
particle momentum becomes random.
After the third collision of the particle, each component of the
momentum becomes random, hence the
particle momentum, for instance $\bs p_a$, becomes completely random.
As a result, we can regard the momentum transfer,
$\bs q = \bs p_a - \bs k_a$, as a random quantity as well.
If we follow the nucleon sequential elastic scattering from the
first collision to the last one, we would see the full randomization
of the particle momentum after each three sequential acts of
scattering.
So, we can look at the process of randomization from the equivalent point
of view: starting from the definite initial momentum $\bs k_0$, the
$n$-th particle gains completely random momentum transfer
$\, {\bs q}_n^{(1)} , \, {\bs q}_n^{(2)} , \ldots , \, {\bs q}_n^{(M)}$
after each three sequential collisions in the chain of $3M$ collisions,
see (\ref{scattering-chain}).

At the same time, the nucleon momentum transfer undergoes the even
faster randomization in the inelastic collisions. Indeed, let us
consider, for example, the process of creation of $N_\pi$ pions in
the nucleon-nucleon reaction: $N+N \to N+N+N_\pi\cdot \pi$. The
process leads to randomization of $[3(2+N_\pi)-4]$ degrees of
freedom, or $d_p$ degrees of freedom per particle becomes random,
where $d_p=3-4/(2+N_\pi)$ and we assume that all particles are on
the mass shell. For instance, if $N_\pi=2$  (see the physical
diagram of the collision depicted in Fig.~\ref{fig:inelastic}), then
two components of the momentum of every particle after a reaction
becomes random, i.e. $d_p=2$ in this process. If $N_\pi \gg 1$, the
number of the random degrees of freedom per particle achieves its
maximal value $d_p=3$. Thus, we see that, in the inelastic
collisions, the randomization of the nucleon momentum transfer
attributed to one physical diagram goes faster than that in the
elastic scattering.

To estimate the {\it effective number of collisions}, $M$, we analyze
all nucleon collisions (physical diagrams), $N_{\rm coll}$, and
obtain a total sum of the random degrees of freedom, $d_{\rm
tot}$, gained by the particle during all reactions before
freeze-out, i.e. we need to know
$d_{\rm tot}=\sum_{i=1}^{N_{\rm coll}} d_p^{(i)}$.
Then, the effective number of collisions is
determined as $M=d_{\rm tot}/3$. So, we assume a randomization of
the momentum transfer, $\bs q$, after every effective collision;
the number of effective collisions, $M$, is determined by the
conditions of a particular experiment.

%%%%%%%%%%%%%%%%%%%%%%%%%%%%%%%%%%%%%%%%%%%%%%%%%%%%%%%%%%%%%%%%%%%%%%
\subsection{Many-particle distribution function for $M$-th ensemble}
%%%%%%%%%%%%%%%%%%%%%%%%%%%%%%%%%%%%%%%%%%%%%%%%%%%%%%%%%%%%%%%%%%%%%%

We would like to determine the density distribution
function
in the momentum space, $f_{2N}(E_{\rm tot};{\bs k}_1,\ldots , {\bs k}_{2N} )$,
which describes $2N$ nucleons in the final
state, after $M$ effective collisions which are experienced by every
nucleon before freeze-out (we name this group of nucleons as the
$M$-th ensemble).
All consideration is carried out in the c.m.s. of two identical
colliding nuclei, hence the initial total momentum of the system
$A+B$ is equal to zero. Let us write down the density distribution
function in the form
%4
\begin{equation}
f_{2N}(E_{\rm tot}; {\bs k}_1,\ldots ,{\bs k}_{2N}) \,
= \, \frac 1{\Omega_{2N}(E_{\mathrm{tot}})} \,
\widetilde{f}_{2N}(E_{\rm tot}; {\bs k}_1,\ldots ,{\bs k}_{2N})  \, ,
\label{A1}
\end{equation}
where we normalize the density distribution function in such a way
that simultaneously determines the density of states in the system:
%9
\begin{equation}
\Omega_{2N}(E_{\mathrm{tot}})
=  \int  d{\tilde k}_1 \ldots  d{\tilde k}_{2N}\,
\widetilde{f}_{2N}(E_{\rm tot}; {\bs k}_1,\ldots ,{\bs k}_{2N} ) \,.
\label{2c}
\end{equation}
The measure of integration in the single-particle phase space
looks like (in units of $\hbar $)
$d{\tilde{k}}_{n}=V\frac{d^{3}k_{n}}{(2\pi )^{3}}\,.$

The unnormalized distribution function
$\widetilde{f}_{2N}(E_{\rm tot};{\bs k}_1,\ldots , {\bs k}_{2N})$
is determined in a two-fold way: first, we follow all collisions of a
particular nucleon by integration with respect to all nucleon random
momentum transfer, and, second, we fix the total energy
%and the total momentum
of the $2N$-nucleon system after freeze-out in a
microcanonical-like way:
$E_{\mathrm{tot}}=\sum_{n=1}^{2N}\omega (\boldsymbol{k}_n)$.
Then, it reads
%4
\begin{equation} \begin{array}{l}\dis
\widetilde{f}_{2N}(E_{\rm tot};{\bs k}_1,\ldots , {\bs k}_{2N} )
=
\delta \! \left( E_{\rm tot} - \sum\limits_{n=1}^{2N} \omega(\bs
k_n) \right) \vspace{4mm}
\\ \dis  \times
\intl \frac{dP_1}{V} \ldots \frac{dP_{2N}}{V} \,
\prod\limits_{n=1}^N \left[  \left(2 \pi \right)^3 \delta^3 \!
\left( {\bs k}_n- {\bs k}_0 - \sum\limits_{m=1}^M {\bs
q}_n^{(m)}\right) \right]
\prod\limits_{n'=N+1}^{2N} \left[ \left(2
\pi \right)^3 \delta^3 \left( {\bs k}_{n'} + {\bs k}_0 -
\sum\limits_{m=1}^M {\bs q}_{n'}^{(m)}\right) \right] \, ,
\end{array}\label{A}
\end{equation}
where $V$ is the volume of the system in the coordinate space.
Here, we assume the independence of the sequential scatterings, which
results in the independence of the particular momentum transfer
(see for details \cite{anch-2008-1}). Hence, the element of the
probability to accept a particular chain of
the momentum transfer by the $n$-th particle in the series of $M$
collisions reads
%5
\begin{equation}
dP_n \equiv \prod_{m=1}^M \, J_m\left({\bs q}_n^{(m)}\right)\,
\frac{d^3q_n^{(m)}}{(2\pi)^3} \, , \label{0}
\end{equation}
where the distribution of the momentum transfer in the $m$-th
collision is characterized by the presence of the form-factor
$J_m({\bs q})$ with
%5
\begin{equation}
\int \frac{d^3q}{(2\pi)^3} \, J_m({\bs q})=1
 \, . \label{ff-normalization}
\end{equation}
In what follows, for the sake of simplicity, we assume the
independence of $J_m(\bs q)$ on the collision number, i.e.
$J_m \left(\bs q \right) \, \rightarrow \ \big \langle J_m
\left(\bs q \right) \big \rangle_{\rm collisions} \,
= \, J \left(\bs q \right)$\,.
Hence, we adopt an approximation where just one form-factor $J(\bs q)$
characterizes a distribution of the momentum-transfer in a series
of collisions which are experienced by a nucleon during its
traveling through the fireball.

With making use of this approximation, we make a step in the
evaluation of the  $2N$-distribution function (\ref{A}).
If one represents
$\delta $-functions in (\ref{A}) in terms of the Fourier integrals
with respect to the auxiliary variables $\boldsymbol{a}_n$,
$\boldsymbol{b}_n$ and then with making allowance for definition
of the integration measure (\ref{0}), the unnormalized density
distribution (\ref{A}) can be written down in the form
%12
\begin{eqnarray}
\widetilde{f}_{2N}(E_{\rm tot};{\bs k}_1,\ldots  ,{\bs k}_{2N} )
&=&
\delta\left( E_{\rm tot}{-} \suml_{n=1}^{2N} \omega (\bs k_n) \right)
\prod_{n=1}^N \Biggl[ \int \frac{d^3a_n}{V}\, e^{-i{\bs a}_n\cdot
({\bs k}_n - {\bs k}_0) } \big[ J(\bs a_n) \big]^M
    \Biggr]
\nonumber \\
&& \hspace{26mm} \times \prod_{n=N+1}^{2N} \Biggr[  \int
\frac{d^3b_n}{V}\, e^{-i{\bs b}_n\cdot ({\bs k}_n + {\bs k}_0) }
\big[ J(\bs b_n) \big]^M   \Biggr] \, ,
\label{A3}
\end{eqnarray}
where
$J(\bs r)=\int \frac{d^3q}{(2\pi)^3} \, J(\bs q) \,
e^{i{\bs q \cdot \bs r} }$
is the Fourier component of the momentum-transfer form-factor
which corresponds to the single nucleon scattering.

We introduce the ``multi-scattering form-factor''
($M$-scattering form-factor)
%15
\begin{equation}
I_M(\bs Q) \equiv \, \frac 1V \, \int d^3r \, e^{-i \bs Q \cdot
\bs r } \, \big[ J( {\bs r} )\big]^M
\, . \label{defi}
\end{equation}
Then we can rewrite the unnormalized distribution function (\ref{A3})
as
%12
\begin{eqnarray}
\widetilde{f}_{2N}(E_{\rm tot};{\bs k}_1,\ldots  ,{\bs k}_{2N} )
&=&
\delta\left( E_{\rm tot}{-} \suml_{n=1}^{2N} \omega (\bs k_n) \right)
\prod_{n=1}^N      I_M({\bs k}_n - {\bs k}_0) \,
\prod_{n'=N+1}^{2N} I_M({\bs k}_{n'} + {\bs k}_0) \, .
\label{A4}
\end{eqnarray}

To obtain the partition function, we make the Laplace
transformation of the density-of-states function (\ref{2c}) with
respect to the variable $E_{\mathrm{tot}}$ (the total energy of the
nucleon system after freeze-out).
As a result, we obtain the partition function of the canonical
ensemble
%8
\begin{equation}
Z_{2N}(\beta)
=
\intl_{E_{\rm min}}^\infty  dE_{\rm tot}\, e^{-\beta E_{\rm tot} } \,
\Omega_{2N}(E_{\mathrm{tot}}) \,.
\label{z2N}
\end{equation}
If we define the unnormalized $2N$-particle distribution function
of the canonical ensemble as
%9
\begin{equation}
\widetilde {\mathbb{F}} _{2N} (\beta ; \bs k_1, \ldots, \bs k_{2N})
=
\intl_{E_{\rm min}}^\infty \, dE_{\rm tot} \, e^{-\beta E_{\rm tot} } \,
\widetilde{f}_{2N}(E_{\rm tot};{\bs k}_1, \ldots ,{\bs k}_{2N} ) \, ,
\label{Ftilde1}
\end{equation}
we can write the partition function as the integral
%10
\begin{equation}
Z_{2N}(\beta)  =  \int d{\tilde k}_1 \ldots d{\tilde k}_{2N} \,
\widetilde {\mathbb{F}}_{2N} (\beta ; \bs k_1, \ldots, \bs k_{2N})
\,. \label{a-12}
\end{equation}
It is obvious that the partition function $Z_{2N}(\beta)$ plays the
role of the
normalization constant with respect to the $2N$-particle distribution
function $\widetilde {\mathbb{F}}_{2N}$ defined in (\ref{Ftilde1}).
Next, we determine the $2N$-particle distribution function in the
canonical ensemble as
%11
\begin{equation}
{\mathbb{F}}_{2N}\left(\beta;{\bs k}_1,\ldots ,
{\bs k}_{2N}\right)
=
\frac 1{Z_{2N}(\beta)} \, \widetilde {\mathbb{F}}_{2N}(\beta; {\bs k}_1,\ldots ,
{\bs k}_{2N} )   \, .
\label{dfce}
\end{equation}
Taking Eq.~(\ref{A4}) into account, for the distribution functions
$\widetilde {\mathbb{F}}_{2N}(\beta;{\bs k}_1,\ldots ,{\bs k}_{2N} )$
from (\ref{Ftilde1}), we obtain
%16
\begin{equation}
\widetilde {\mathbb{F}}_{2N}(\beta;{\bs k}_1,\ldots ,{\bs k}_{2N}
)= \prod_{n=1}^N  \Big[ e^{ -\beta \omega(\bs k_n) } I_M({\bs k}_n
-{\bs k}_0)
 \Big] \,
\prod_{n'=N+1}^{2N} \Big[ e^{ -\beta \omega(\bs k_{n'}) } \,
I_M({\bs k}_{n'} + {\bs k}_0)  \Big] \, . \label{Ftilde}
\end{equation}
So, we can write the distribution function which characterizes the
$2N$-nucleon system where each particle experiences $M$ effective
collisions ($M$-th ensemble) in a factorized form
%17
\begin{equation}
{\mathbb{F}}_{2N}(\beta,{\bs k}_1,\ldots ,{\bs k}_{2N} )
=
\prod_{n=1}^N  f_a(\bs k_n) \, \prod_{n=N+1}^{2N}  f_b(\bs k_n)
\, , \label{Ffact}
\end{equation}
where
%18
\begin{equation}
f_{a(b)}(\bs k)
=
\frac 1{z_{a(b)}(\beta)} \, e^{ -\beta \omega(\bs k) } \,
I_M(\bs k \mp {\bs k}_0)  \quad {\rm with}
\quad z_{a(b)}(\beta)
=
V\, \int \frac{d^3k}{(2\pi)^3} \, e^{ -\beta \omega(\bs k) } \,
I_M(\bs k \mp{\bs k}_0) \, ,
 \label{1DF}
\end{equation}
are the single-particle distribution function and the
single-particle partition function, respectively, attributed to
nucleus ``A'' for subindex $a$ or to nucleus ``B'' for
subindex $b$.

%%%%%%%%%%%%%%%%%%%%%%%%%%%%%%%%%%%%%%%%%%%%%%%%%%%%%%%%%%%%%%%%%%%%%%
\subsubsection{Saddle-point approximation}
\label{section:saddle-point-approximation}
%%%%%%%%%%%%%%%%%%%%%%%%%%%%%%%%%%%%%%%%%%%%%%%%%%%%%%%%%%%%%%%%%%%%%%

For large enough collision numbers $M$, we can calculate the
integral in (\ref{defi}) within the saddle-point method.
To do this, we first represent the integral in the form
$I_M(\bs Q) = \, \frac 1V \, \int d^3r \,
\exp{\left[-i \bs Q \cdot \bs r + M \ln{J(\bs r)} \right] } $\,.
Then we expand the exponent $\ln{J({\bs r})}$ into a series at
the point $\bs r=0$ up to the second order.
We take into account that $J(\bs r)\big|_{\bs r=0} = 1$ (see Eq.
(\ref{ff-normalization})).
If the form-factor depends on the modulus of the momentum
transfer $J(|\bs q|)$, the expansion of the logarithm $\ln{J({\bs r})}$
looks like
\begin{equation}
\ln{J(r)} \approx \, - \frac 16 \, \langle q^2 \rangle \, r^2 \,,
\qquad {\rm where} \qquad
\langle q^2 \rangle = \int \frac{d^3q}{(2\pi)^3}
{\bs q}^2 \, J(|\bs q|)
\label{1Dexp}
\end{equation}
with $q=|\bs q|$ and $r=|\bs r|$.
Here, we take into account that $\partial_i J(\bs r)\big|_{\bs r=0} = 0$,
which is valid when $J(-\bs q)=J(\bs q)$.
Finally, we obtain integral (\ref{defi}) in the approximate
form
\begin{equation}
I_M(Q) \approx \frac 1V \left( \frac{6\pi} { M \langle q^2
\rangle } \right)^{3/2}
\exp{ \left(- \frac{3\, Q^2}{2 M \langle q^2\rangle} \right)} \ ,
\label{i-largeM}
\end{equation}
where $Q=|\bs Q|$.

So, the $M$-scattering form-factor $I_M(Q)$ for a large number of the
effective collisions $M$ can be approximately represented as the
normal distribution $\propto \exp{\left(-Q^2/2\sigma^2\right)}$ with
the variance $\sigma=\sqrt{M \langle q^2\rangle/3}$\,, which is obviously
a reflection of the central limit theorem.

Following approximation (\ref{i-largeM}), the single-particle
distribution functions (\ref{1DF}) take the form
%18
\begin{equation}
f_{a(b)}(\bs k)
=
\frac 1{z_M(\beta)} \, e^{ -\beta \omega(\bs k)
- \frac{3\, (\bs k \mp {\bs k}_0)^2}{2 M \langle q^2\rangle}} \,
  \quad {\rm with}
\quad z_M(\beta)
=
\int \frac{d^3k}{(2\pi)^3} \, e^{ -\beta \omega(\bs k)
- \frac{3\, (\bs k - {\bs k}_0)^2}{2 M \langle q^2\rangle}} \,,
\label{1DFapprox}
\end{equation}
where we skip the common factors in $f_{a(b)}(\bs k)$ and in
$z_M(\beta)$, respectively.
In the limit case $M \to \infty$, the dependence on the initial momentum
$\pm \bs k_0$ is washed out, and both single-particle distributions
$f_a(\bs k)$ and $f_b(\bs k)$ take the same ``thermal'' limit:
\begin{equation}
f_{a(b)}(\bs k) \  \rightarrow \ f_{\rm therm}(\bs k)
=
\frac 1{z_{\rm therm}(\beta)} \, e^{ -\beta \omega(\bs k) }  \,,
\quad z_{\rm therm}(\beta) = \int \frac{d^3k}{(2\pi)^3} \,
e^{ -\beta \omega(\bs k) } \, .
 \label{DFtherm}
\end{equation}

In the non-relativistic case, i.e. when $\omega(\bs k)=\bs k^2/2m + m$,
one can rewrite
distributions (\ref{1DFapprox}) as the J\"{u}ttner ones \cite{juttner}
(see \cite{groot})
%18
\begin{equation}
f_{a(b)}(\bs k)
=
\frac 1{z_M(\beta)} \,
\exp{ \left[-\frac{(\bs k \mp m\bs v_{\rm h})^2}{2mT_{\rm eff}} \right]} \,
  \quad {\rm with} \quad
z_M(\beta)
=
\int \frac{d^3k}{(2\pi)^3} \,
\exp{ \left[-\frac{(\bs k - m\bs v_{\rm h})^2}{2mT_{\rm eff}} \right]} \,.
\label{1DFjuttner}
\end{equation}
Here, we define the collective (hydrodynamical) velocity $\bs v_{\rm h}$
as
\begin{equation}
\bs v_{\rm h} = \left(\frac 1{1+\frac \beta{\beta_{\rm coll}(M)}}
\right) \, \frac {\bs k_0}{m} \qquad {\rm with} \qquad \beta_{\rm
coll}(M) \equiv \frac {3m}{M\langle q^2\rangle} \,, \label{T1}
\end{equation}
and the effective temperature as
%18
\begin{equation}
T_{\rm eff} = \frac 1{\beta + \beta_{\rm coll}(M) }
\,. \label{T2}
\end{equation}
The quantity $\beta_{\rm coll}$ can be put in correspondence to the
``collision'' temperature
$T_{\rm coll} = 1/\beta_{\rm coll}
= \frac 23 M\langle \omega(\bs q)\rangle$,
where
$\langle \omega(\bs q)\rangle=\langle q^2\rangle/2m$ is the mean
energy transfer in one nucleon collision.
We see that every collision ensemble has its own effective temperature
and own collective velocity.
Indeed, with increase in the number of collisions $M$, the effective
temperature increases and the collective velocity is going down.
These quantities have the following limits:
%18
\begin{equation}
 (M=0) \left| \begin{array}{rcccl}
\dis \frac{k_0}{\omega(k_0)}  & \geq & v_{\rm h} & \geq & 0  \vspace{2mm}\\
0  & \leq  & \dis T_{\rm eff}  & \leq  & T \end{array} \right|
 (M \to \infty)
\label{limits}
\end{equation}
where $\omega(k_0)=\sqrt{m^2+k_0^2}$, $T=1/\beta$, and we marked
conventionally the nucleons which do not take part in any collisions
(spectators) by $M=0$. Here, the left limit corresponds to the case
$M=0$, and the right limit corresponds to $M\to \infty$. Hence, the
every step during the increase in the number of collisions results
in the redistibution of the energy accumulated initially in the
longitudinal movement, i.e. $v_{\rm h}=k_0/\omega(k_0)$ and $T_{\rm
eff}=0$ before first collision.
The energy passes partially to the transverse degrees of freedom,
which increases the isotropization and the temperature of the system
and decreases, in turn, the collective velocity.
When the number of collisions is large enough, i.e. when
$M\to \infty$, we come to the limits: $\bs v_{\rm h} \to 0$ and
$T_{\rm eff} \to T $. These conclusions are valid, of course,
for relativistic energies as well.

Meanwhile, when the number of collisions is finite we see that, just
for the non-relativistic dispersion law, the obtained distribution
(\ref{1DFapprox}) can be regarded as a locally thermal one
(the J\"{u}ttner distribution function).

%%%%%%%%%%%%%%%%%%%%%%%%%%%%%%%%%%%%%%%%%%%%%%%%%%%%%%%%%%%%%%%%%%%%%%
\subsubsection{Single-particle spectrum for $M$-th ensemble}
%%%%%%%%%%%%%%%%%%%%%%%%%%%%%%%%%%%%%%%%%%%%%%%%%%%%%%%%%%%%%%%%%%%%%%

We can construct a ``two-source'' single-particle
distribution function $ f(\bs k_a,\bs k_b)= f_a(\bs k_a)\, f_b(\bs
k_b)$. Then the averaging of the two-source random quantity
$W(\bs k_a,\bs k_b)$ gives
%16
\begin{equation}
\langle W \rangle
=
\int \frac{d^3k_a}{(2\pi)^3}\, \frac{d^3k_b}{(2\pi)^3} \, W(\bs
k_a,\bs k_b)\, f(\bs k_a,\bs k_b) \, .
\label{tsa}
\end{equation}
If we take $ W(\bs k_a,\bs k_b) = \frac 12 \left[ \delta^3({\bs
p}-{\bs k}_a)+ \delta^3({\bs p}-{\bs k}_b) \right]$, we obtain,
after the averaging, the single-particle spectrum which is attributed
to the system where all particles experienced $M$ collisions before
freeze-out:
%17
\begin{equation}
D_M(\bs p)
= \left(\frac 1{2N} \frac{d^3 N}{d^3p} \right)_M = \frac 12 \,
e^{-\beta \omega({\bs p})} \left[ \frac{I_M(\bs p - {\bs
k}_0)}{z_a(\beta)} + \frac{I_M(\bs p + {\bs k}_0)}{ z_b(\beta)}
\right],
\label{sps}   % sps = single-particle spectrum
\end{equation}
where $I_M(\bs Q)$ is defined in (\ref{defi}).
Hence, with accounting for notations (\ref{1DF}), the distribution
functions $D_M(\bs p)$ can be represented also as
%18
\begin{equation}
D_M(\bs p)
= \frac 12 \, \left[ \,
f_a(\bs p) + f_b(\bs p) \, \right] \, .
\label{sps2}   % sps = single-particle spectrum
\end{equation}
For large collision numbers $M$, the distribution function $D_M(\bs p)$
reads
%19
\begin{equation}
D_M(\bs p)
\approx
\frac {e^{-\beta \omega({\bs p})}}{2z_M(\beta)} \,
 \left[ e^{ - \frac{3(\bs p - \bs k_0)^2}{2M \langle q^2 \rangle}}
 +  e^{ - \frac{3(\bs p + {\bs k}_0)^2} {2M \langle q^2 \rangle}}
 \right] \quad {\rm with} \quad
z_M(\beta) = \int \frac{d^3k}{(2\pi)^3}
e^{-\beta \omega(\bs k) - \frac{3(\bs k - \bs k_0)^2}{2M
\langle q^2 \rangle}} \, .
\label{dm}
\end{equation}
It is evidently seen that the spectrum has two items which
can be attributed to the first and to the second colliding nuclei,
respectively. In accordance with this structure, it can be named
a ``two-source single-particle spectrum''. It is clear also
that
this picture is just the effective one: after a nucleon-nucleon collision, we
cannot distinguish nucleons, and thus we cannot say anything about their
origination from a particular nucleus.

In what follows, we use the distributions $D_M(\bs p)$ from
(\ref{sps}) to describe a nucleon spectrum arising in relativistic
nucleus-nucleus collisions.

%%%%%%%%%%%%%%%%%%%%%%%%%%%%%%%%%%%%%%%%%%%%%%%%%%%%%%%%%%%%%%%%%%%%%%
\subsection{Nucleon Rapidity Distribution and Transverse Spectrum}
%%%%%%%%%%%%%%%%%%%%%%%%%%%%%%%%%%%%%%%%%%%%%%%%%%%%%%%%%%%%%%%%%%%%%%

To obtain the transverse mass and rapidity distributions, we pass
to new variables $(p_x,p_y,p_z) \to (\phi, m_\bot, y)$:
$m_\bot=(m^2+{\bs p}_\bot^2)^{1/2}$, \quad ${\bs p}_{\bot }^{2} =
p_{x}^{2}+p_{y}^{2}$, \quad $\dis \tanh{y}= \frac{p_z}{\omega(\bs p)} $,
then
$d^3p=d\phi \, \omega_p \, m_\bot \, dm_\bot \, dy$, where $\phi$ is
the azimuth angle.
In accordance with (\ref{model2}), the double differential cross-section
reads (for the sake of simplicity, we consider here the most central
collisions which possess the azimuth symmetry)
%20
\begin{equation}
\frac{d^{\,2} N}{ m_\bot dm_\bot dy} = 2\pi\, m_\bot \cosh y
\left[ \ \sum_{M=1}^{M_{\rm max}}  C(M) \, D_M(m_\bot,y)
+ \, C_{\rm therm} \, D_{\rm therm}(m_\bot,y) \, \right] \, .
\label{tr-total}
\end{equation}
We can define the distribution functions in the rapidity space as
%21
\begin{equation}
\Phi_M(y)  \equiv 2\pi \cosh{y} \int_{m}^\infty
dm_{\perp} \, m_{\perp}^2 \, D_M(m_\bot,y) \quad
{\rm with} \quad \int\, dy \, \Phi_M(y) = 1 \, .
\label{phiM}
\end{equation}
Then the rapidity distribution looks like
%22
\begin{equation}
\frac{dN}{dy} \, = \, \sum_{M=1}^{M_{\rm max}} C(M) \, \Phi_M(y) +
C_{\rm therm} \, \Phi_{\rm therm}(y) \, ,
\label{y-fit}
\end{equation}
where
\begin{equation}
\Phi_{\rm therm}(y) = \frac{2\pi \cosh{y}}{z_{\rm therm}(\beta) } \,
\frac{ e^{-x} }{ \beta^3 \cosh^3{y} } (x^2+2x+2)
%\quad {\rm with} \quad x=m\beta \cosh{y}
\label{phitherm}
\end{equation}
with $x=m\beta \cosh{y}$ and
$z_{\rm therm}(\beta) = 4\pi \frac{m^2}{\beta} \, K_2(m\beta) $.

%%%%%%%%%%%%%%%%%%%%%% Fig1 %%%%%%%%%%%%%%%%%%%%%%%%%%%%%%%%%%%%%%%%%%
\begin{figure}[t]
\includegraphics[width=0.58\textwidth]{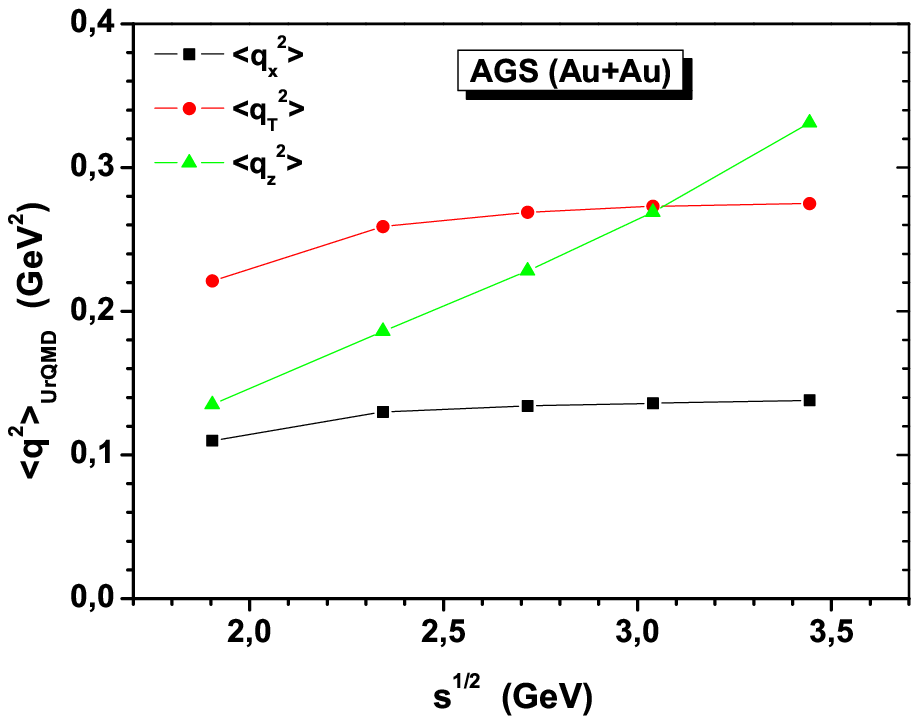}
\hfill
\includegraphics[width=0.48\textwidth]{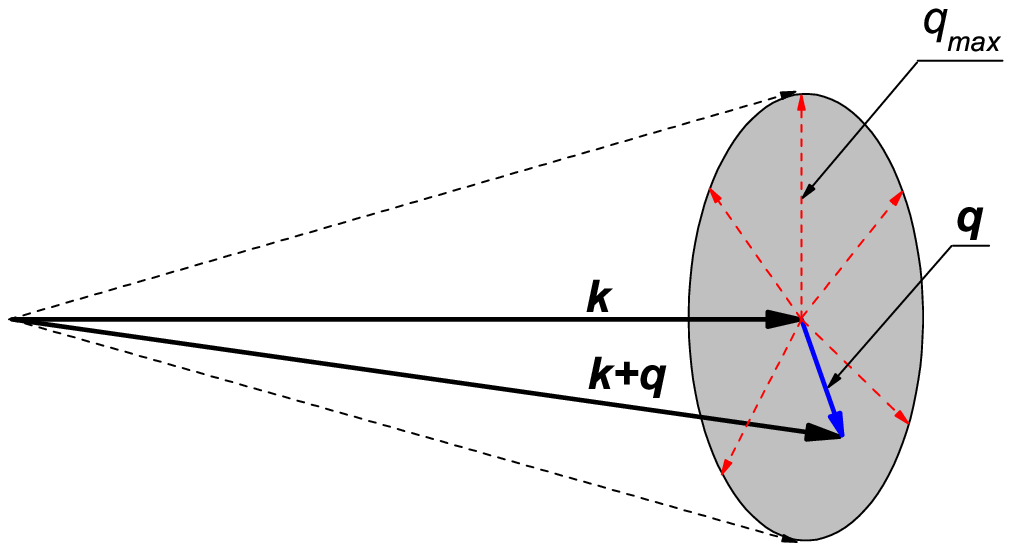}
\\
\parbox[t]{0.48\textwidth}{  \caption
{The mean value of the nucleon momentum transfer squared as a function
of the collision energy obtained in the framework of the UrQMD
\cite{urqmd1,urqmd2} simulations for the most central collisions
at AGS energies. }
\label{fig:meanQS}} \hfill
\parbox[t]{0.48\textwidth}{  \caption
{Nucleon momentum transformation as a result of the two-particle
collision; $\bs k$ is the initial momentum, $\bs q$ is the momentum
transfer, $\bs k + \bs q$ is the momentum after the reaction,
$q_{\rm max}$ is the maximally allowed modulus of the
momentum transfer.} \label{fig:conus}}
%  \end{minipage}
\end{figure}
%%%%%%%%%%%%%%%%%%%%%%%%%%%%%%%%%%%%%%%%%%%%%%%%%%%%%%%%%%%%%%%%%%%%%%

%%%%%%%%%%%%%%%%%%%%%%%%%%%%%%%%%%%%%%%%%%%%%%%%%%%%%%%%%%%%%%%%%%%%%%
\section{Toy model}
%%%%%%%%%%%%%%%%%%%%%%%%%%%%%%%%%%%%%%%%%%%%%%%%%%%%%%%%%%%%%%%%%%%%%%

We would like to obtain the explicit results in the framework of the
proposed approach. We note that the crucial quantity of our approach
is the form-factor $J(\bs q)$ which can be considered as the
probability density (see (\ref{ff-normalization})) to find the
nucleon momentum transfer $\bs q$ in a particular nucleon collision.
Then the mean value of the quantity which depends on the momentum
transfer, for instance $f(\bs q)$, is obtained as $\langle f \rangle
=\int d^3q/(2\pi)^3\, f(\bs q)\, J(\bs q)$. To estimate the behavior
of the form-factor $J(\bs q)$, we made evaluation of the mean values
of $q_x^2$, $q_\bot^2$, and $q_z^2$ in the framework of the UrQMD
\cite{urqmd1,urqmd2}. The results of the evaluation are depicted in
Fig.~\ref{fig:meanQS}. It is evident that, for the central
collisions, the equality $\langle q_x^2 \rangle =\langle
q_y^2\rangle $ should be valid; we see this, indeed, in
Fig.~\ref{fig:meanQS}, $\langle q_\bot^2 \rangle \approx 2 \langle
q_x^2 \rangle $. It is seen that, for AGS energies, the mean values
$\langle q_z^2 \rangle $ and $\langle q_x^2 \rangle$ approximately
equal one another or of the same order. We take this as a basis to
formulate a model: let the form-factor $J(\bs q)$ be chosen as a
homogeneous distribution of the momentum transfer in the sphere of
finite radius $q_{\rm max}$ (see Fig.~\ref{fig:conus}),
\begin{equation}
J({\bs q})
=
\frac{(2\pi)^3}{V_q} \, \theta(q_{\rm max}-|\bs q \,|) \, , \quad
{\rm where} \quad V_q= \frac 43 \, \pi \, q_{\rm max}^3  \quad
{\rm and} \quad \int \frac{d^3q}{(2\pi)^3} \,J({\bs q})=1 \, .
\label{toym}
\end{equation}
As a matter of fact, an application of this toy model can be regarded
just as an approximate description of the nucleon distributions at AGS
energies.

We calculated the Fourier transform of form-factor (\ref{toym}) in
the explicit way and obtained
%10
\begin{equation}
J(x)
=\frac 3{x^3} \, \left( \, \sin {x} -x\, \cos{x}  \, \right)  \,,
\label{j-explicit}
\end{equation}
where $ x \equiv\ r q_{\mathrm{max}}$ and $r=|\bs r|$.
So, the multiscattering form-factor $I_M(\bs Q)$, see (\ref{defi}),
reads now as
%10
\begin{equation}
I_M(Q)
=
\frac{4\pi}{V Q} \int_0^\infty  dr \, r\, \sin{ (Q\, r) }
\left[ \frac 3{x^3} \, \left(\sin {x} - x\, \cos{x}\right)\right]^M
  \,,
\label{i-explicit}
\end{equation}
where $Q=|\bs Q|$, $ x = r q_{\mathrm{max}}$.

With the help of these functions, we can evaluate the partial
distribution $D_M(\bs p)$ from (\ref{sps}) as
%17
\begin{equation}
D_M(m_\bot,y)
= \frac 1{2z(\beta)} \,
e^{-\beta m_\bot \cosh y} \big[ \, I_M(|\bs p - {\bs k}_0|) +
I_M(|\bs p + {\bs k}_0|) \, \big] \, ,
\label{tmsps}
\end{equation}
where $\omega({\bs p}) = m_\bot \cosh y$, $p_z = m_\bot \sinh y$,
$|\bs p \mp {\bs k}_0| = \sqrt{m_\bot^2-m^2 +
(m_\bot \sinh y \mp k_{0z})^2}$ \,
and
%22
\begin{equation}
z(\beta)
=
V \int \frac{d^3p}{(2\pi)^3} \, e^{ -\beta \omega(\bs p) } \,
I_M(|\bs p - {\bs k}_0|) \, .
\label{s-ppf}
\end{equation}
Next, we use these partial distribution functions directly in
(\ref{tr-total}) to evaluate the double differential (transverse
mass) spectrum and the functions $\Phi_M(y)$ from (\ref{phiM}) which
will be used then as the partial rapidity distributions in
(\ref{y-fit}).

In the framework of the toy model, we obtain that the parameter of the
model, $q_{\mathrm max}$, can be related to the mean value of the
momentum transfer squared
%18
\begin{equation}
\left\langle \bs q^2 \right\rangle = \, \frac 35 \, q_{\mathrm max}^2
 \, .
\label{menaq}
\end{equation}
Taking into account $\left\langle \bs q^2 \right\rangle$ evaluated
in UrQMD \cite{urqmd1,urqmd2} for AGS energies (see
Fig.~\ref{fig:meanQS}), we obtain the following estimation from
(\ref{menaq}): $q_{\mathrm max} \approx 800$~MeV.

In the framework of the toy model for large collision numbers $M$,
the distribution functions $D_M(m_\bot,y)$ reads
%19
\begin{equation}
D_M(m_\bot,y)
\approx
\frac {e^{-\beta m_\bot \cosh y}}{2z(\beta)} \,
 \left[ e^{ - \frac{5(\bs p - \bs k_0)^2}{2M q_{\mathrm max}^2}}
 +  e^{ - \frac{5(\bs p + {\bs k}_0)^2} {2M q_{\mathrm max}^2}}
 \right] \quad {\rm with} \quad
z(\beta) = \int \frac{d^3p}{(2\pi)^3}
e^{-\beta \omega(\bs p) - \frac{5(\bs p - \bs k_0)^2}{2M q_{\mathrm max}^2}}
\, ,
\label{dm-tm}  % dm-tm = D_M in toy model
\end{equation}
where
$|\bs p - {\bs k}_0|$
and
$|\bs p + {\bs k}_0|$ are defined in the same way as in (\ref{tmsps}).

%%%%%%%%%%%%%%%%%%%%%%%%%%%%%%%%%%%%%%%%%%%%%%%%%%%%%%%%%%%%%%%%%%%%%
\subsection{Extraction of the physical information from
experimental data}
%%%%%%%%%%%%%%%%%%%%%%%%%%%%%%%%%%%%%%%%%%%%%%%%%%%%%%%%%%%%%%%%%%%%%%

By making use of the rapidity distribution
(\ref{y-fit}), we fit the experimental data on the rapidity
distribution of net protons which were measured at the AGS (E802
Collaboration) \cite{E802-PRC-v60-064901-1999}.
The proton data is remarkable in that sense that we know exactly the
initial momentum, $k_{0z}$, of every nucleon.

We note that, in the case of a small number of experimental points
$N_{\rm exp}$, the set of functions $\Phi_M(y)$ is overcomplete
($N_{\rm exp} < M_{\rm max}+1$). To choose a unique configuration of
the variable parameters $C(M)$, we use the maximum entropy method
\cite{soroko,papoulis}. Namely, we define the information entropy
%22
\begin{equation}
\sigma \, = \, - \sum_{M=1}^{M_{\rm max}+1} \overline C(M) \,
\ln{\overline C(M)} \, ,
\label{ientropy}
\end{equation}
where $\overline C(M)=C(M)/N_p$, $N_p$ is the total number of
protons, and, for the unification of notations, we adopt $C(M_{\rm
max}+1)\equiv C_{\rm therm}$. Our goal is to find the maximum of the
information entropy $\sigma$. Meanwhile, entropy (\ref{ientropy}) is
supplemented by $(N_{\rm exp}+1)$ constraints:
\[
\sum_{M=1}^{M_{\rm max}+1} \overline C(M) = 1 \,, \qquad
I_i^{\rm (exp)} = \sum_{M=1}^{M_{\rm max}+1} C(M)\, \Phi_M(y_i) \,,
\]
where $I_i^{\rm (exp)}$ is an experimental data value of the
distribution $dN/dy$ taken at the rapidity point $y_i$,
$i=1,2, \ldots, N_{\rm exp}$ and
$\Phi_{M_{\rm max}+1}(y) \equiv \Phi_{\rm therm}(y)$.
Using the method of the Lagrangian multipliers, we reformulate the
problem in the following way: it is necessary to find the maximum of the
expression
\[
L= - \sum_{M=1}^{M_{\rm max}+1} \overline C(M) \, \ln{\overline
C(M)} + \mu \left[\, \sum_{M=1}^{M_{\rm max}+1} \overline C(M) - 1
\right] + \sum_{i=1}^{N_{\rm exp}} \lambda_i \left[ I_i^{\rm (exp)}
- N_p \, \sum_{M=1}^{M_{\rm max}+1} \overline C(M)\, \Phi_M(y_i)
\right] \,, \] where $\mu$ and $\lambda_i$ are the $(N_{\rm exp}+1)$
Lagrangian multipliers. Derivatives of this expression with respect
to all unknown coefficients give a set of $\left(M_{\rm
max}+1+N_{\rm exp}+1\right)$ equations which we solve using the
variation method: first, we exclude $\mu$ and all $\overline C(M)$
from the system of equations and left with a reduced system of
$N_{\rm exp}$ transcendental equations for the unknown Lagrangian
multipliers $\lambda_i$, where $i=1,2, \ldots, N_{\rm exp}$. The
latter reduced system of equations is solved by variations of the
$\lambda_i$. So, we seek out the minimum of the expression
%22
\begin{equation}
\chi^2(\lambda_1,\lambda_2, \ldots ,\lambda_{N_{\rm exp}})=
\sum_{i=1}^{N_{\rm exp}}  \left[ I_i^{\rm (exp)} -
N_p \, \sum_{M=1}^{M_{\rm max}+1} \overline C(M)\, \Phi_M(y_i) \right]^2
\,, \label{hi2}
\end{equation}
where $\overline C(M)=X_M/\left(\sum_{M=1}^{M_{\rm max}+1}X_M
\right)$ with $X_M=\exp{\left[N_p\sum_{i=1}^{N_{\rm exp}} \lambda_i
\Phi_M(y_i)\right]}$. Actually, Eq.~(\ref{hi2}) is nothing more as a
total $\chi^2$, with a help of which we find a theoretical curve of
the closest fit to $dN/dy$ experimental data. At the same time, it
is easy to rewrite (\ref{hi2}) with accounting for normalization
with respect to the experimental error bars at every rapidity point
$y_i$.

\begin{figure}[h]
\includegraphics[width=0.48\textwidth]{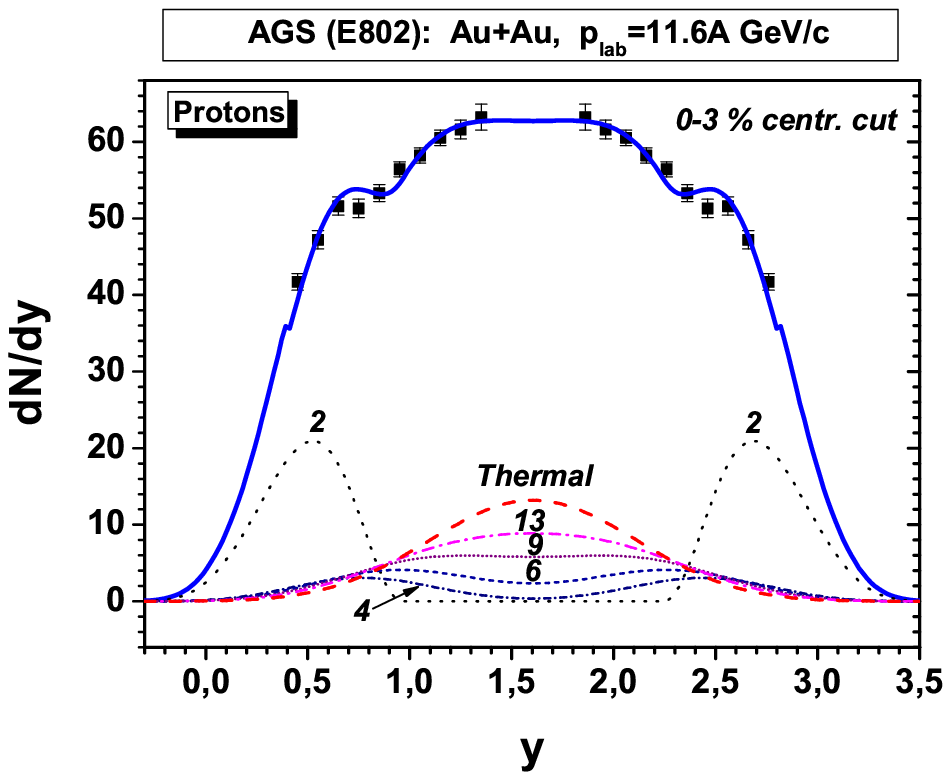} \hfill
\includegraphics[width=0.48\textwidth]{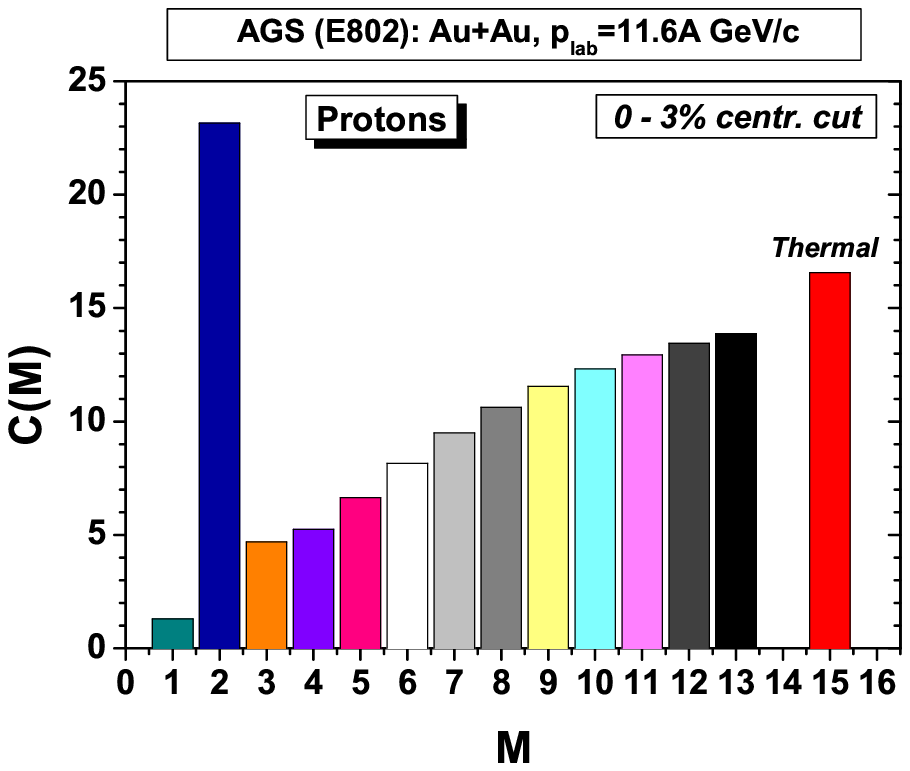}
\parbox[t]{0.48\textwidth}{\caption
{The result of the fit (thick solid curve)
to experimental data
  \cite{E802-PRC-v60-064901-1999}
  on rapidity distribution.
  Broken curves marked by numbers $M$ represent the partial
  contributions from every collision ensemble,
  $C(M)\, \Phi_M(y)$, the thermal contribution is represented by
  the central Gaussian-like dashed curve
  .} \label{fig:2a}} \hfill
\parbox[t]{0.48\textwidth}{\caption{
The values of the coefficients $C(M)$ which were obtained as the
result of a fit of the $dN/dy$ data \cite{E802-PRC-v60-064901-1999}.
The histogram of the coefficient $C_{\rm therm}$ is putted on the
place $M=15$ just for a convenience of comparison.  }
\label{fig:histogram}}
\end{figure}

The fit was carried out with a help of the program MINUIT, and the
variable parameters are the Lagrangian multipliers $\lambda_i$. The
coefficients $C(M)$ and $C_{\rm therm}$ are evaluated then through
the obtained multipliers $\lambda_i$. The slope parameter $\beta$
was first extracted from the double differential yield for protons
with the use of the thermal distribution. All evaluations are
carried out in the c.m.s. of colliding nuclei. The obtained
theoretical curves are depicted in Fig.~\ref{fig:2a}. The solid
thick curve represents the result of the fit, total $\chi^2 = 13.3$.
The broken curves  marked by the numbers $M$ represent the partial
contributions from every ensemble, $C(M) \, \varphi_M(y)$. The
thermal contribution is represented by the central bell-like dashed
curve. The obtained parameters for $T= 1/\beta = 280$~MeV are shown
in Table 1. In the form of histograms, these coefficients are
depicted in Fig.~\ref{fig:histogram}.
{\small
%%%%%%%%%%%%%%%%%%%%%%%%%%%%%%%%%%%%%%%%%%%%%%%%%%%%%%%%%%%%%%%%%%%%%%
\noindent\begin{tabular}{c c c c c c c c c c c c c c c}
&&&&&&&&&&&&& Table 1 \\
\hline%
   \multicolumn{1}{|c}{$C(1)$} &
   \multicolumn{1}{|c}{$C(2)$} & \multicolumn{1}{|c}{$C(3)$}
 & \multicolumn{1}{|c}{$C(4)$} & \multicolumn{1}{|c}{$C(5)$}
 & \multicolumn{1}{|c}{$C(6)$} & \multicolumn{1}{|c}{$C(7)$}
 & \multicolumn{1}{|c}{$C(8)$} & \multicolumn{1}{|c}{$C(9)$}
 &   \multicolumn{1}{|c}{$C(10)$}
 & \multicolumn{1}{|c}{$C(11)$} & \multicolumn{1}{|c}{$C(12)$}
 & \multicolumn{1}{|c}{$C(13)$} & \multicolumn{1}{|c|}{ $C_{\rm therm}$ }
 \\%
\hline%
   \multicolumn{1}{|c}{1.3} &
   \multicolumn{1}{|c}{23.2} & \multicolumn{1}{|c}{4.7}
 & \multicolumn{1}{|c} {5.3} & \multicolumn{1}{|c} {6.6}
 & \multicolumn{1}{|c} {8.2} & \multicolumn{1}{|c} {9.5}
 & \multicolumn{1}{|c}{10.6}  & \multicolumn{1}{|c}{11.6}
 & \multicolumn{1}{|c}{12.3}
 & \multicolumn{1}{|c}{12.9} & \multicolumn{1}{|c}{13.5}
 & \multicolumn{1}{|c}{13.9} & \multicolumn{1}{|c|}{ 16.6  }
 \\%
\hline
\end{tabular}
%%%%%%%%%%%%%%%%%%%%%%%%%%%%%%%%%%%%%%%%%%%%%%%%%%%%%%%%%%%%%%%%%%%%%%
}
\medskip

%
%%%%%%%%%%%%%%%%%%%%% Figs %%%%%%%%%%%%%%%%%%%%%%%%%%%%%%%%%%%%%%%%%%%
%%%%%%%%%%%%%%%%%%%%%%%%%%%%%%%%%%%%%%%%%%%%%%%%%%%%%%%%%%%%%%%%%%%%%%

\noindent So, each coefficient $C(M)$ tells us how many protons
undergo $M$ effective collisions or which is the population of each
collision ensemble. For instance, the ensemble of protons which
participated just in two effective collisions, $M=2$, consists of
$23$ protons, i.e. $C(2)=23.2$. It is worth to note that this
ensemble is maximally populated in comparison to others. What is
very important, we learn from this expansion that $C_{\rm therm}=
16.6$. This means that approximately seventeen protons come from a
thermal source, what makes up $11$\% (eleven percent) of all
participated protons. The maximal number of collisions, $M_{\rm
max}=13$ was determined from the UrQMD evaluation
\cite{urqmd1,urqmd2} as the maximum number of effective proton
collisions which is obtained by averaging over the number of
Monte-Carlo events. Actually, the histograms depicted in
Fig.~\ref{fig:histogram} represent a tomography picture with respect
to the number of collisions.

\begin{figure}[h]
\includegraphics[width=0.49\textwidth]{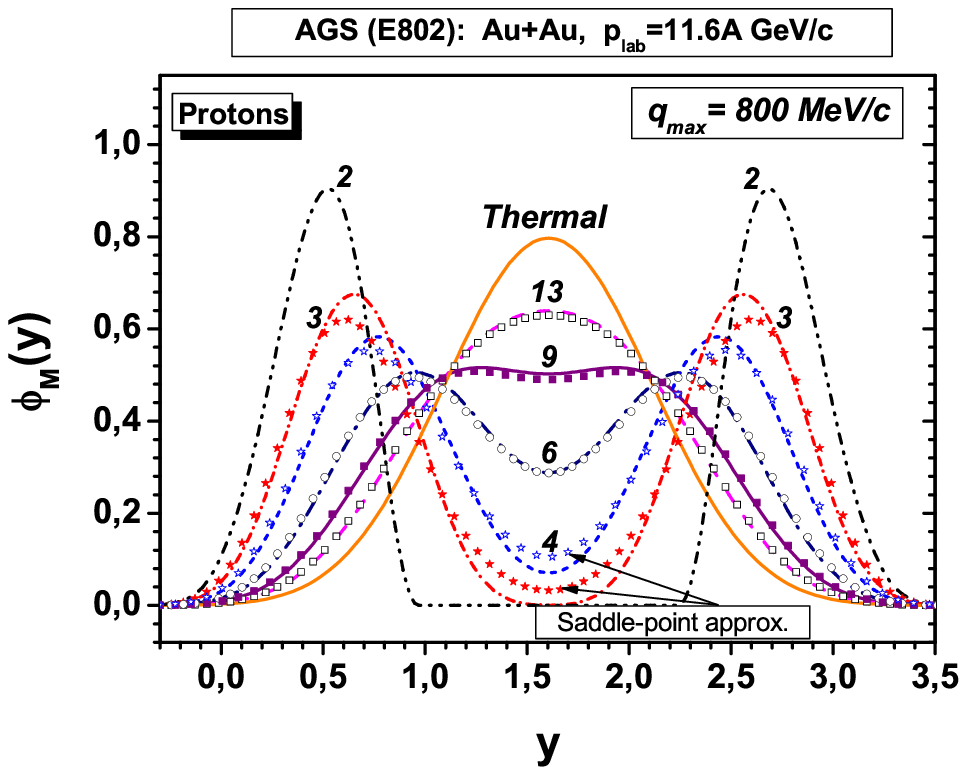} \hfill
\includegraphics[width=0.48\textwidth]{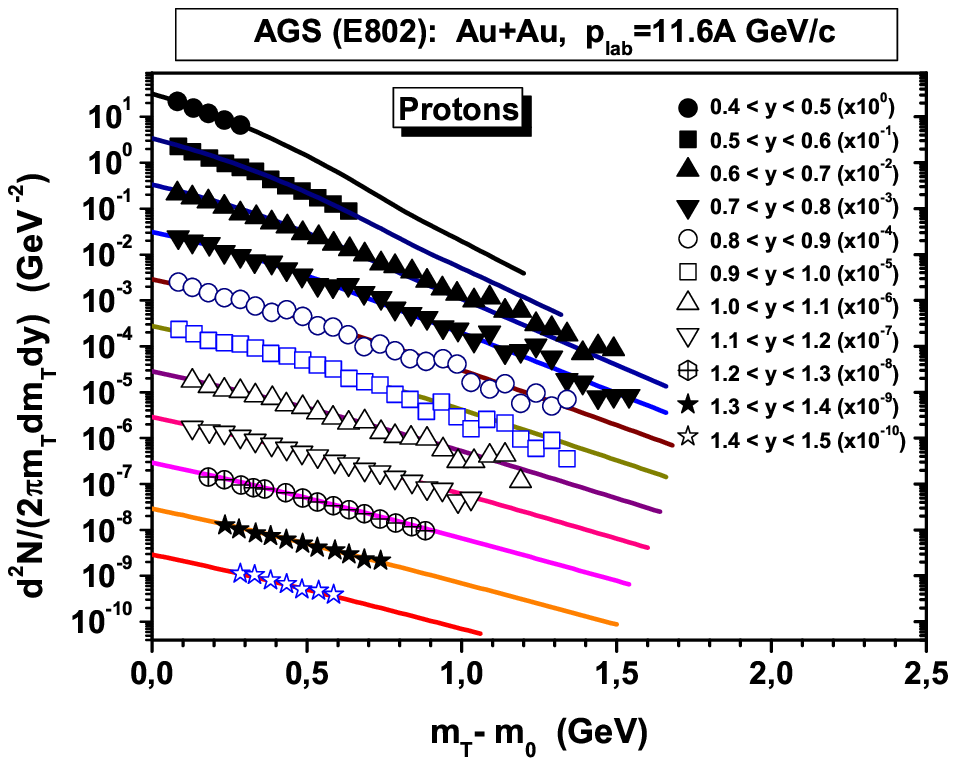}
\\
\parbox[t]{0.48\textwidth}{\caption
{  The curves marked by numbers $M$ represent the functions
  $\Phi_M(y)$ (see (\ref{phiM})), the thermal distribution
  $\Phi_{\rm therm}(y)$ is   represented by the central Gaussian-like
  curve.
  The scatter curves represent the same $\Phi_M(y)$ evaluated in the
  saddle-point approximation. }
\label{fig:phiM}} \hfill
\parbox[t]{0.48\textwidth}{\caption{
Solid curves  represent the evaluation of the
  $m_{\perp}$-spectra obtained in accordance
  with Eq. (\ref{tr-total}), where we use the same values of the
  coefficients $C(M)$ which were obtained as the result of a fit of
  the $dN/dy$ data (see Fig.~\ref{fig:2a}).
  Experimental points are from \cite{E802-PRC-v60-064901-1999}.}
\label{fig:2b}}
\end{figure}

The partial distribution functions in the rapidity space,
$\Phi_M(y)$ and $\Phi_{\rm therm}(y)$ (see (\ref{phiM}) and
(\ref{phitherm})), are depicted in Fig.~\ref{fig:phiM}. One can say
that the total rapidity distribution $dN/dy$ from (\ref{y-fit}) is
nothing more as an expansion over the set of these $M_{\rm max}+1$
functions. In Fig.~\ref{fig:phiM}, we depicted also, as the scatter
curves, the functions $\Phi_M(y)$ evaluated in the framework of the
saddle-point approximation (see (\ref{1DFapprox})). Starting from
$M=4$, these curves, as seen, differ very slightly from the
distribution evaluated in a rigorous way.

So, we see that the top of partial rapidity distributions which
corresponds to the rapidity point $y_{\rm top}$ is shifted to the
central rapidity region with increase in the number of effective
collisions $M$ (in Fig.~\ref{fig:phiM}, the number $M$ marks every
particular distribution). This means that the collective velocity
$v_{\rm h}\approx \tanh{y_{\rm top}}$ of the distribution
$\Phi_M(y)$ decreases with increase in $M$. Evidently, the energy
which was initially accumulated in the longitudinal degrees of
freedom is converted to the particle creation and to the excitation
of the transverse degrees of freedom with decrease in the
longitudinal velocity $v_{\rm h}$, i.e. after every collision. The
latter results in the collision-by-collision increase of the
isotropization and the effective temperature. These two conclusions
are in full correspondence with the analysis made in Section
\ref{section:saddle-point-approximation}.

By making use of the obtained coefficients $C_M$ and $C_{\rm
therm}$, which are listed in Table~1 (see histograms of the
coefficients in Fig.~\ref{fig:histogram}), we evaluated the double
differential cross-section in accordance with (\ref{tr-total}). The
results of this evaluation for different rapidity windows is
depicted in Fig.~\ref{fig:2b}. A very good description of the
experimental data \cite{E802-PRC-v60-064901-1999} for all rapidity
windows is seen. Hence, our model satisfactory describes the
$m_{\perp}$-spectra.

We assume that the thermal source has absolutely different nature of
origination, i.e. it cannot be created just due to the hadron
reactions of nucleons which result in the randomization and the
subsequent isotropization of the nucleon momentum. The thermal
source can emerge as a result of the appearance of many other (not
hadron) degrees of freedom. We know just one candidate to this role,
it is the quark-gluon plasma, for instance, its creation can occur
in collisions of nucleons at high energies in the presence of a
dense medium. A many-parton system, which emerges in the collision,
consists of a large number of gluons and  quarks. All momenta of
quarks and gluons can be regarded from the very beginning as random
ones, and the thermalization of the system occurs during a time span
$\tau_{\rm therm}=0.6$~fm/c \cite{Gyulassy-0709.1716}. Hence, the
protons which come from the thermal source indicate the presence of
the QGP in the fireball, and we can determine a nucleon power of the
QGP, $P_{\rm qgp}$, by the ratio of the number of protons coming
from the thermal source, $C_{\rm therm}$, and total number of
participated protons: $P_{\rm qgp}\equiv C_{\rm therm} /N_p$. For
instance, it turns out that the nucleon power of the QGP equals
eleven percent, $P_{\rm qgp}\approx 11\%$, in $Au+Au$ collisions
\cite{E802-PRC-v60-064901-1999} (0-3\% centrality).

%%%%%%%%%%%%%%%%%%%%% Figs %%%%%%%%%%%%%%%%%%%%%%%%%%%%%%%%%%%%%%%%%%%
%%%%%%%%%%%%%%%%%%%%%%%%%%%%%%%%%%%%%%%%%%%%%%%%%%%%%%%%%%%%%%%%%%%%%%

%%%%%%%%%%%%%%%%%%%%%%%%%%%%%%%%%%%%%%%%%%%%%%%%%%%%%%%%%%%%%%%%%%%%%%
\section{Summary and discussion}
%%%%%%%%%%%%%%%%%%%%%%%%%%%%%%%%%%%%%%%%%%%%%%%%%%%%%%%%%%%%%%%%%%%%%%

The starting assumption of our approach is given in
Eq.~(\ref{model1}): we propose to separate the registered nucleons
by their origination into two groups. The first group consists of
the net nucleons after the multiple rescattering and the second
group consists of the nucleons created by the QGP. We consider the
latter as a source of the fully thermalized nucleons which is in the
rest in the c.m.s. of colliding nuclei. Our model is applied at AGS
and low SPS energies, where the creation of nucleon-antinucleon
pairs is comparatively small.

The second assumption consists in full independence of the momentum
transfer in the chain of sequential collisions of a nucleon:
\[ J \left(\bs q^{(1)},\ldots,\bs q^{(M)}\right) =
\prodl_{m=1}^{M} J_m \left(\bs q^{(m)} \right) \,, \]
where $J\left(\bs q^{(1)},\ldots,\bs q^{(M)}\right)$ is the
probability density to find the set $\bs q^{(1)},\ldots,\bs q^{(M)}$
of the momentum transfer during $M$ collisions of the nucleon in the
fireball; $J_m(\bs q)$ is the probability density  to find the
momentum transfer $\bs q$ in the $m$-th collision. Additionally, we
assume that the scattering conditions do not depend on the collision
number:
\[
J_m \left(\bs q \right) \, \rightarrow \ \big \langle J_m
\left(\bs q \right) \big \rangle_{\rm collisions} \,
= \, J \left(\bs q \right) \,.
\]

The essence of the third assumption is a partition of all nucleons,
which took part in rescattering processes, into collisions ensembles
in accordance with the effective collision number $M$, see
Eq.~(\ref{model1}). The physical basis of this assumption is a
spatial finiteness of the many-nucleon system.

So, we parametrize the time axis by the number of particle's
effective collisions $M=\langle \nu \rangle t$, where $\langle \nu
\rangle$ is the mean frequency of collisions and $t$ is the time
interval. If we considers the dependence of the distribution
function (\ref{sps}) or (\ref{dm}) on the variable $M$, we can
regard this distribution as a nonequilibrium one ($M \propto$ time).
As we see from (\ref{dm}), the increase of $M$ effectively mimics an
approach to the Boltzmann distribution (for the sake of simplicity,
we consider no special statistics). If the number of collisions $M$
is fixed, this means that the process of thermalization stopped at
the time moment when the particles had experienced just $M$
collisions and, at this moment, they were frozen out; all this
results in a partial thermalization of the subsystem which we name
as the $M$-th collision ensemble. So, the level of thermalization
and isotropization depends on the number of effective collisions,
$M$, which is determined, first, by the lifetime of the system or,
more precisely, by the number of physical collisions of every
particle and, second, by the level of randomization of the momentum
transfer in every physical collision.

Actually, in the framework of the proposed model, we obtain a
physical picture which looks like a discrete ``fireball model'',
where every fireball can be associated with a particular collision
ensemble. In accordance with the fireball model
\cite{becattini-cleymans-2007}, the rapidity axis is populated with
thermalized fireballs following a distribution $\rho(y_{\rm h})$
which is taken as the Gaussian one. (Here, the rapidity $y_{\rm h}$
determines the collective (hydrodynamical) velocity $v_{\rm
h}=\tanh{y_{\rm h}}$ of the particles attributed to the particular
fireball.) The essential difference which occurs in our model,
except the discretization of the fireball set, is that ``our''
fireballs (collision ensembles) are not yet the fully thermalized
systems. The degree of thermalization of every collision ensemble is
frozen on some stage of evolution toward a full thermalization (the
collective velocity of the particles which belong to the particular
ensemble can be evaluated approximately as $v_{\rm h}\approx
\tanh{y_{\rm top}}$, where $y_{\rm top}$ determines the top of the
$\Phi_M(y)$ distribution in the rapidity space, see
Fig.~\ref{fig:phiM}).

Properly, the results of the proposed model crucially depend on the
single-particle form-factor $J(\bs q)$ which is nothing more as the
probability density to find a particular value $\bs q$ of the
momentum transfer in one collision. As a trial evaluation (toy
model), we take the form-factor as a homogeneous restricted
distribution of the momentum transfer, $\dis J(\bs q)= \frac{(2
\pi)^3}{V_q} \,\theta(q_{\rm max} - |\bs q|)$, where $\dis V_q =
\frac {4 \pi}3 q_{\rm max}^3$ (see Fig.~\ref{fig:conus}).

The maximum number of collisions (reactions) $M_{\rm max}$
is assumed to be finite and determined by the nuclear number $A$,
initial energy, and centrality. With the help of the UrQMD
transport model \cite{urqmd1,urqmd2}, it was found that, under AGS
(Au+Au, 11.6A GeV/c, 0-3\% centrality) conditions
\cite{E802-PRC-v60-064901-1999}, $M_{\rm max}=13$. Using the
thermal distribution, we extract the slope parameter from the
experimental data on the proton $m_{\perp}$-spectra, $T=280$~MeV.

We made fit of the experimental data \cite{E802-PRC-v60-064901-1999}
on the rapidity distribution of net protons and obtained the
collection of coefficients $C(M)$ (see Table~1 and
Fig.~\ref{fig:histogram}) which are nothing more as the absolute
number of protons in every collision ensemble, i.e.
$N_p=\sum_{M=1}^{M_{\rm max}} C(M)+ C_{\rm therm}$, where $N_p$ is
the total number of protons.

The knowledge of the number of protons, $C_{\rm therm}$, which
come from the QGP gives us a possibility to evaluate the
``nucleon power'' of the QGP, $P_{\rm QGP}^{(N)}$, created in a
particular experiment on the nucleus-nucleus collision. We find
that, under the AGS conditions \cite{E802-PRC-v60-064901-1999} (a
centrality of 0-3\%), $P_{\rm QGP}^{(N)}\approx 11\%$. So, in the
framework of the proposed criterion, it could be claimed that the
QGP (as a nucleon source) was created not only at SPS energies
\cite{heinz2000}, but it was also created in the central
collisions at AGS energies.

Meanwhile, if we do not use the thermal contribution in the
partial expansion (\ref{y-fit}) for the description of the
particular experimental data \cite{E802-PRC-v60-064901-1999}, we
obtain as well a good description (the same $\chi^2$) of both the
rapidity distribution and $m_{\perp}$-spectra. So, we cannot
resolve unambiguously the problem about the presence of the
thermal source. In fact, to overcome the problem, we need a more
detailed experimental information for the central rapidity region.

Solid curves in Fig.~\ref{fig:2b}  represent the results of
calculations of the $m_{\perp}$-spectra obtained in accordance with
Eq. (\ref{tr-total}), where we use the same values of the
coefficients $C(M)$ (see Table~1) which were obtained as the result
of a fit of the $dN/dy$ data. It is seen from the results depicted
in Figs.~\ref{fig:2a} and \ref{fig:phiM} that the main contribution
to the distribution for small rapidity values (in the laboratory
system) is given by collision ensembles with small $M$. For
instance, for the particular experiment
\cite{E802-PRC-v60-064901-1999}, the partial distribution for $M=2$
in Eq. (\ref{y-fit}) determines the distribution in the rapidity
windows $0.4 \le y \le 0.6$. We note that, in this rapidity domain,
the thermal contribution is practically zero. So, we can conclude
that the proposed model satisfactorily describes the
$m_{\perp}$-spectra even in those regions, where the contribution
from a pure thermal source is absent and the spectrum is determined
only by the rescattering of net nucleons.

All this leaves us with the continued challenge of applying the
model to other experiments and problems.

%%%%%%%%%%%%%%%%%%%%%%%%%%%%%%%%%%%%%%%%%%%%%%%%%%%%%%%%%%%%%%%%%%%%%%
\section*{Acknowledgements}
%%%%%%%%%%%%%%%%%%%%%%%%%%%%%%%%%%%%%%%%%%%%%%%%%%%%%%%%%%%%%%%%%%%%%%

D.A. is grateful to A.~Rebhan, S.~Mr\'owczy\'nski, and K.~Tuchin for
useful and encouraging discussions. The authors are grateful to
R.~Lednicky for reading the manuscript and for reasonable remarks.
D.A. was partially supported by the program \textquotedblleft
Fundamental properties of physical systems under extreme
conditions\textquotedblright\ (Division of the Physics and Astronomy
of the NAS of Ukraine).

\end{document}